\documentclass[sigconf, nonacm]{acmart}

\AtBeginDocument{%
  \providecommand\BibTeX{{%
    \normalfont B\kern-0.5em{\scshape i\kern-0.25em b}\kern-0.8em\TeX}}}

\usepackage{siunitx}
\usepackage{tabularx}
\usepackage{balance}
\begin{document}

\title{SuaCode Africa: Teaching Coding Online to Africans using Smartphones}


\author{George Boateng}
\affiliation{%
  \institution{ETH Z{\"u}rich}}
 \affiliation{\institution{Nsesa Foundation}}
\email{gboateng@ethz.ch}

\author{Prince Steven Annor}
\affiliation{\institution{New York University Abu Dhabi}}
\affiliation{\institution{Nsesa Foundation}}
\email{princeannor@gmail.com}

\author{Victor Wumbor-Apin Kumbol}
\affiliation{\institution{Charit\'e Berlin}}
\affiliation{\institution{Nsesa Foundation}}
\email{vkumbol@nsesafoundation.org}

\renewcommand{\shortauthors}{Boateng et al.}

\begin{abstract}
There is a burgeoning trend of smartphone ownership in Africa due to the low costs of Android smartphones and the global increase in social media usage. Building upon previous works that introduced a smartphone-based coding course to secondary and tertiary students in Ghana via an in-person program and an online course, this work introduced Africans in 37 countries to our online smartphone-based course in 2019. Students in this 8-week course read lesson notes, submitted assignments, collaborated with peers, and facilitators in an online forum and completed open and closed-ended surveys after the course. We performed qualitative and quantitative analyses on the data from the course. 

Out of the 709 students that applied, 210 were officially admitted to the course after passing the preliminary assignments. And at the end of the course, 72\% of the 210 students completed the course. Additionally, students’ assignment submissions and self-reports showed an understanding of the programming concepts, with comparable performance between males and females and across educational levels. Also, students mentioned that the lesson notes were easy to understand and they enjoyed the experience of writing code on their smartphones. Moreover, students adequately received help from peers and facilitators in the course forum. Lastly, results of a survey sent to students a year after completing this program showed that they had developed various applications, wrote online tutorials, and learned several tools and technologies. We were successful at introducing coding skills to Africans using smartphones through SuaCode Africa.

\end{abstract}



\maketitle

\section{Introduction}
Africa lags behind the rest of the world in terms of digital literacy. Less than 1\% of African children leave school with basic coding skills \cite{sap2016}. Africa is home to the “largest and youngest workforce in the world” but several “companies and governments are struggling to fill IT-related positions” \cite{sap2016}. 

Yet, there is an interest among Africans to learn to code. This interest is evidenced by several initiatives in Africa to introduce coding with a prominent one being Africa Code Week \cite{acw}. Africa Code Week is organized by SAP and UNESCO in partnership with over 130 organizations across Africa. The goal of Africa Code Week is to “multiply free coding workshops for youth across 37 countries” while “building local trainer capacity through Train-the-Trainer sessions” and “increasing girl participation”. The program has already benefited over 4 million young Africans so far from 2015 to 2019. The success of Africa Code Week illustrates the need and desire of Africa’s young population to learn to code. 

However, the lack of access to equipment such as computers for teaching and learning hampers efforts to teach coding to Africans. Also, Africans without easy access to computers cannot keep practicing coding once the program ends which is key for deepening skills development. This problem has been heightened by the COVID-19 pandemic because most coding initiatives rely on in-person training in centers with the necessary computers and facilitators. However, there is a proliferation of smartphones in Africa. According to research firm Ovum, there will be 929.9 million smartphones in Africa by the year 2021 \cite{matinde2017}. Hence, smartphones provide a unique means of circumventing the problem of lack of access to computers to introduce coding to Africans, giving them the opportunity to practice and build their skills at the tip of their fingers, literally.

In 2017, we developed a smartphone-based coding course in Ghana — a first in the country — that enabled 27 students most of whom did not have computers to learn to code on their phones during an in-person workshop \cite{boateng2018}. The course was very successful with its objectives, but 72.8\% of the 77\% of students that used smartphones  during the course said they did not enjoy the smartphone coding experience \cite{boateng2018}. This feedback was due to reasons like small screen sizes, and crashes of the coding app on phones with lower Android versions. In 2018, we adapted the smartphone-based course for delivery as an online course and we ran a pilot with 30 students in Ghana \cite{boateng2019}. That pilot showed that students could be introduced to coding on their smartphones via an online course, with improved smartphone-coding experience albeit with some challenges such as a low completion rate of 23\% and a need for an automated grading system to reduce the burden on facilitators. In 2019, we built upon that work and scaled the course beyond Ghana and ran SuaCode Africa, introducing 709 Africans in 37 countries to our online smartphone-based course — a first in Africa. We addressed some of the limitations of the previous work by using a selection criteria for students which resulted in higher completion rate (72\%) and we also built an automated grading system which enabled us to scale to hundreds of students. This work is important now more than ever, given the COVID-19 pandemic has highlighted the need for research, development, and evaluation of online education strategies. In this experience report, we describe the program, our evaluation and insights gained.

\section{Related Work}
Various studies have investigated the role of mobile phones and smart devices in education \cite{Jacob2007MobileLI, Aderinoye_Ojokheta_Olojede_2007, Shonola2016TheIO, Ayoub, Kukulska}. Some studies have also showed that students store and access course material from their mobile devices because it improved information retention due to factors like easy accessibility, portability and increased flexibility for self-study \cite{Jacob2007MobileLI, Shonola2016TheIO}. Students are also able to take advantage of the inherent communicative nature of smartphones to collaborate on projects and assignments and also help their peers \cite{Kukulska}. Maleko et al. also showed that Mobile Social Learning Environments encouraged novice students to interact and engage more with their learning \cite{maleko}.  In a survey of university students in Nigeria \cite{Shonola2016TheIO}, being able to learn anywhere and at anytime was one of the reasons cited for the use of mobile devices for learning. 

Furthermore, Elkhateeb et al \cite{elkhateeb2019mobile} developed a mobile learning system "Easy-Edu" for students in an Egyptian university. "Easy-Edu" was evaluated by administering a C\# programming course and the authors reported significantly higher learning performance in students who used the mobile learning system in comparison to those who had traditional face-to-face learning. In this paper, several limitations of earlier mobile learning systems were highlighted. These included the lack of peer discussion and assistance tools for collaboration between students, the need for game-based modules and multi-platform support.

Additionally, the small screen size and keypads available on mobile phones were identified as limitations to their use in programming education by Mbogo et al. \cite{mbogo2014supporting}. In a subsequent study \cite{mbogo2016evaluating}, the authors designed a scaffolded programming environment to teach Java programming on mobile phones. This scaffolded environment was evaluated in an experiment involving  university students in Kenya and South Africa. The study revealed that scaffolding techniques enable learners to attempt and complete more programming tasks with fewer run-time errors.

In our work, we delivered and evaluated a smartphone-based course that taught coding in particular in an online environment. This work differs from those that delivered general courses using mobile devices \cite{Jacob2007MobileLI, Aderinoye_Ojokheta_Olojede_2007, Shonola2016TheIO, Ayoub, Kukulska}. This work also differ from those of Elkhateeb et al. in which they evaluated their mobile learning system with an in-person course (n=100) in one country. Though the context of the course was a programming course, students did not write code on their phones and hence they did not evaluate the smartphone-coding experience \cite{elkhateeb2019mobile}. The work by Mbogo et al. on the other hand evaluated smartphone-based coding in an in-person course in 2 countries \cite{mbogo2016evaluating}. However, our work is the first that delivers and evaluates a smartphone-based coding course in an online environment and at scale — 709 students across Africa (37 countries).
  
\section{Overview of SuaCode Africa}
SuaCode Africa was run as a 2-month online course for high school, university students and recent graduates living in Africa and the diaspora. We put out various advertisements and received applications from 709 Africans (25\% female) in 37 countries with a majority living in Africa, and Nigeria and Ghana being the two most represented countries. These advertisements were published on websites, sent through social media groups and pages, and emailed to educational and non-profit organizations.



All 709 students were conditionally admitted and invited to the course platform for a trial period of 3 weeks. Official admissions were contingent on students making submissions for the first two programming assignments (described later) regardless of the grade scored. They had access to all the course materials and resources during this period. We used these criteria to ensure that only committed participants were admitted into the program. Out of 709 students, we officially admitted 210  applicants comprising individuals at various levels of education: high school (secondary school) students  (11.0\%), university (tertiary) students (41.1\%), university graduates (46.4.0\%), primary school students (1.0\%), and high school graduates (0.5\%). Most of the admitted students were resident in Africa (99.3\%) with the remainder in the diaspora. 
  
\begin{figure}[ht]
  \centering
  \includegraphics[width=0.9\linewidth]{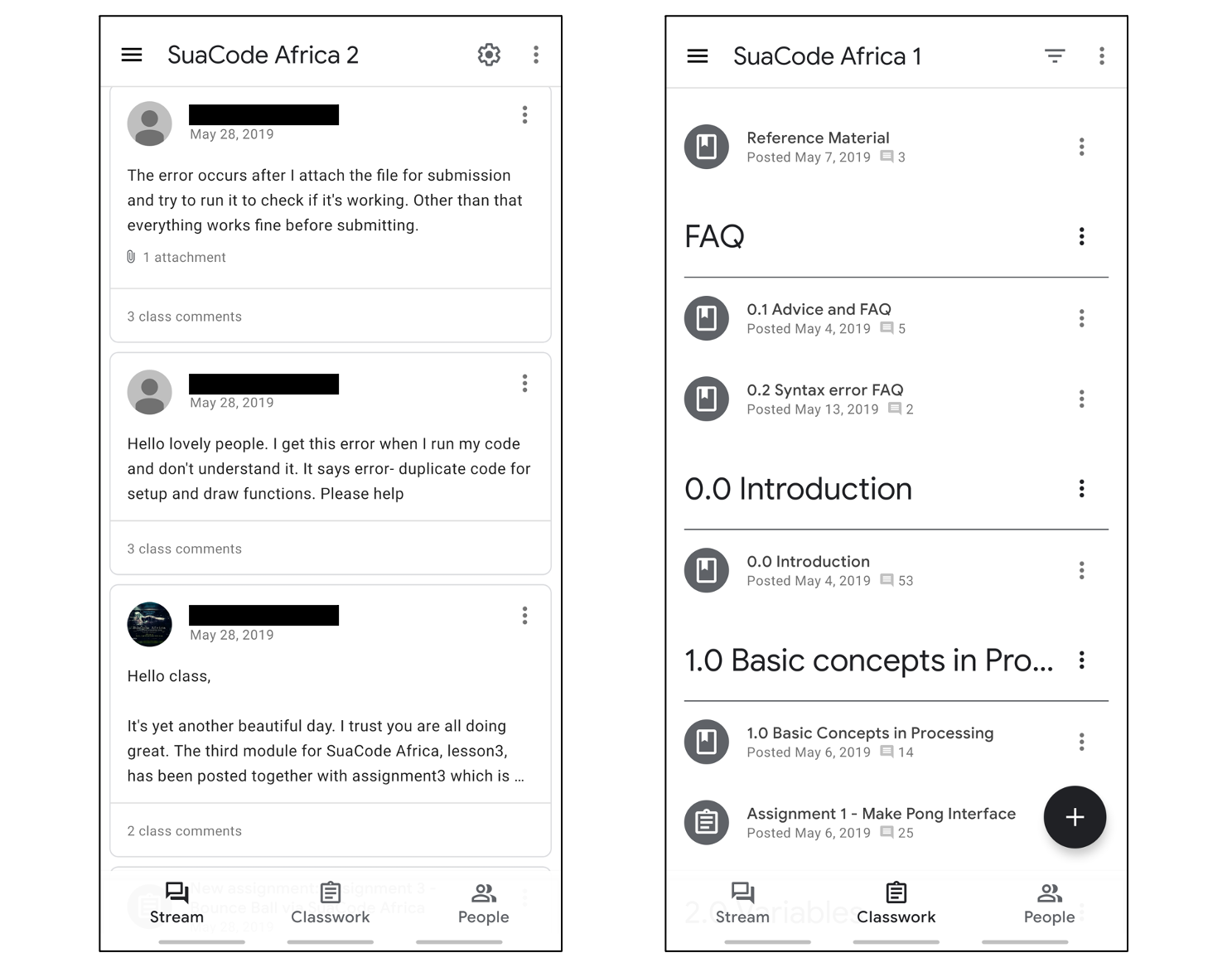}
  \caption{Snapshot of Google Classroom}
  \Description{A virtual classroom}
\label{fig:classroom}
\end{figure}

\begin{figure}[ht]
  \centering
  \includegraphics[width=0.9\linewidth]{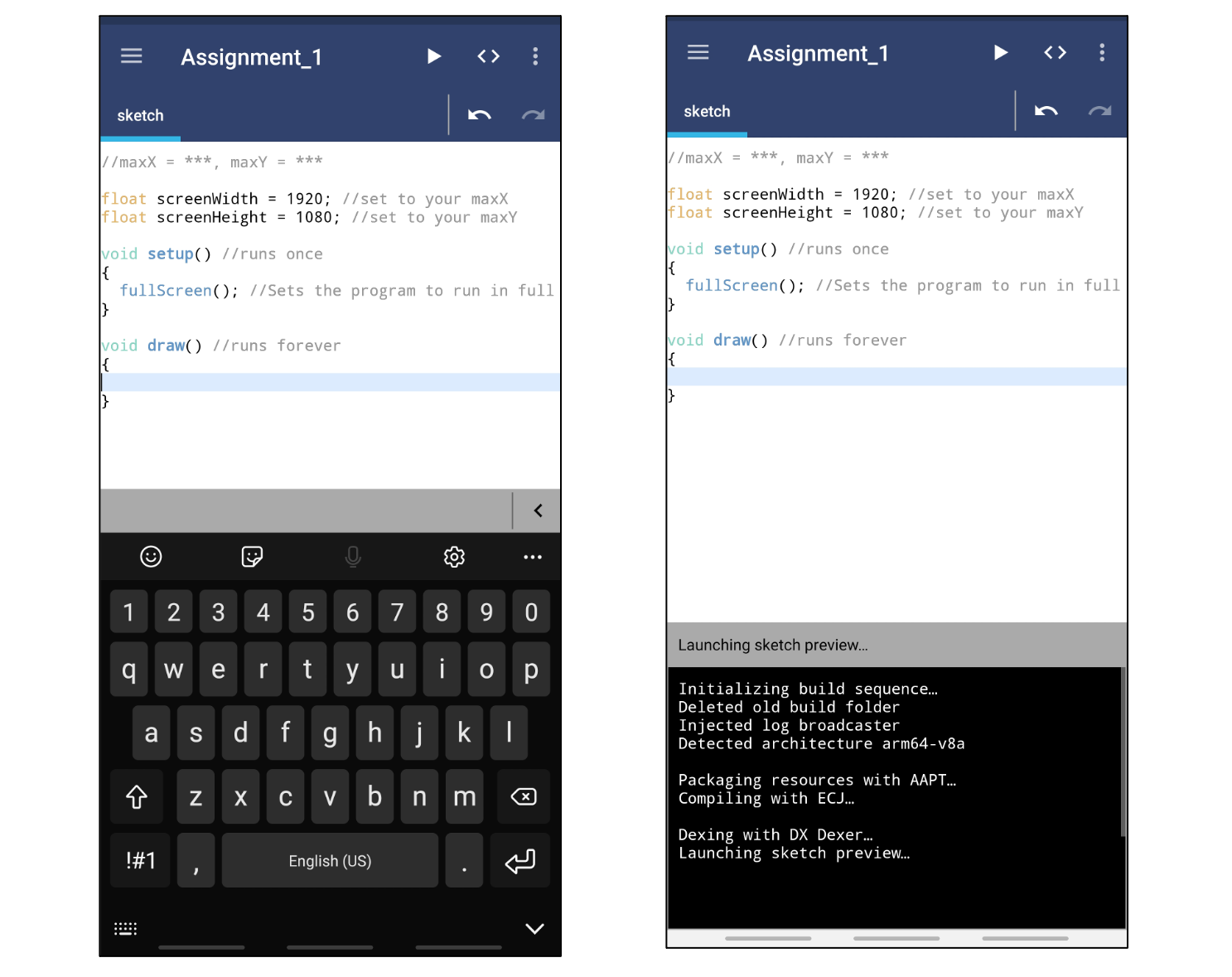}
  \caption{Snapshot of Android Processing Development Environment (APDE) application}
  \Description{An app}
\label{fig:apde}
\end{figure}

Similar to our past work \cite{boateng2019}, we hosted and delivered the course using the learning management system, Google Classroom, which was used to deliver the text-based lesson notes (as Google docs), receive assignment submissions and provided a forum for students to post questions and receive answers (Figure \ref{fig:classroom}). We did not use videos as instruction materials but only lesson notes because Internet data is  expensive in Africa, and more so for our target population, students living in Africa \cite{QuartzAfrica:2019}. Google Classroom was used because the files could be made offline, it was free for teachers and students, and it was streamlined and simple to use on a smartphone. We had 14 facilitators that answered students’ questions on the course forum. These facilitators were at junior developer rank, and were trained with the course materials. The students were sent a link to join Classroom via their emails and were sent the introductory setup document to join Classroom and setup the APDE Android App that was used for writing code in the course (Figure \ref{fig:apde}). Students supported each other with questions and answers, and were guided by facilitators when they faced issues via the forum provided in Classroom. A live frequently asked questions Google document was also accessible to the students via the Classroom platform. 

\begin{figure}[ht]
  \centering
  \includegraphics[width=0.6\linewidth]{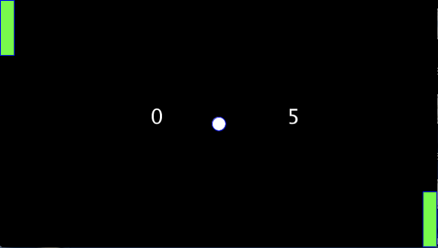}
  \caption{Snapshot of pong game on Android phone}
  \Description{A pong game}
\label{fig:pong}
\end{figure}

We used part 1 of our smartphone-based coding curriculum \cite{boateng2019}. The curriculum uses the Processing programming language, an open-source, Java-based programming language \cite{processing} which was chosen because it enables learning of programming in a fun way since it can be used to easily create visual and interactive programs. Students used the APDE Android app for writing and running their programs \cite{apde}. The curriculum covers four lessons: Basic Concepts in Processing (Lesson 1), Variables (Lesson 2), Conditionals (Lesson 3) and Functions (Lesson 4),  and results in the building of a Pong game  (Figure \ref{fig:pong}) \cite{boateng2018, boateng2019}. This game has 2 paddles, one for each player and a ball. Once the ball starts moving, each player has one goal — to prevent the ball from exiting the vertical wall on their side using the paddles. If the latter happens, the opponent’s score increases. Lesson 1 introduces students to Processing and some basic concepts in graphical programming. Lesson 2 then builds upon lesson 1 and introduces variables and standard coding practices. Lesson 3 introduces conditionals, specifically if-else statements, and lesson 4 introduces functions and best modular coding practices. Each lesson is made available as a Google document, with code example files and has a corresponding assignment that incrementally builds a component of a pong game \cite{suacode-course}. 
  
Assignment 1 entails building the interface of the pong game, assignment 2 makes the ball move by incorporating variables to store states, assignment 3 makes the ball bounce off the game walls by using conditionals, and assignment 4 gets students to write functions to move the paddles and put it all together. The assignments were also available as Google docs. Students received automated notifications when assignment deadlines were close. Additionally, reminder messages were posted on the forum by facilitators. 


 Students were able to make the lessons and assignments available offline once they were released. The lesson notes and assignments were made available on specific dates throughout the 8-week period along with specified hard deadlines. This cohort-based model is quite different from most online coding courses such as those on Coursera, for example, which tend to be self-paced with students progressing through the entire course material at their own pace. It rather resembles regular in-person courses that tend to have hard deadlines. We used this approach to ensure that students committed to completing the course within the defined period. Cohort-based learning fulfills students’ need for affiliation, support, interaction, and teamwork \cite{maher2005}. The progress of students was tracked with their assignment submissions. Some students from similar countries also formed private groups on platforms like WhatsApp to support each other. 

The assignments were mostly graded by AutoGrad, an automated grading system \cite{autograd} that we developed to address one of the shortcomings of our previous work in which the manual grading of assignments was burdensome. Since AutoGrad was still in development during the cohort, some edge cases and the last assignment were manually graded. AutoGrad downloads the student’s assignment from Google’s servers and provides students with feedback on their assignment code submissions with pointers to missing aspects of the assignment’s specifications. AutoGrad also gives feedback on the usage of comments and indenting. To grade assignments, we ran both static and dynamic code analysis and compared the student’s code with the specifications for the assignments. With static analysis, we first parsed the student’s code file and employed regular expressions checks. An example of static analysis is as follows. In assignment one, students have to draw a ball at the center of the screen. AutoGrad in this case checks if a circle has been drawn via searching for the function call for a circle and where it was drawn based on its parameters. With dynamic analysis, the student’s code gets wrapped into a class definition, an object gets instantiated to run the code file, and then various checks are performed by manipulating various parameters and checking how other parameters change. An example of dynamic analysis is checking if, for example, the ball bounces off the top and bottom walls correctly. To do this, we run the student’s code, make the ball move towards the top and bottom walls, and check if various parameters such as the ball’s movement direction changes as expected. 


\section{Evaluation and Discussion}
Our criteria for completion was that students submitted assignments 1 through 4, and scored a cumulative grade of at least 50 out of the 80 total. At the end of the course, 151 out of the 210 admitted students completed and received certificates, resulting in a completion rate of 72\%, an improvement over that of the pilot of SuaCode (23\%)\cite{boateng2018} and much higher than the average for massive open online courses (20\%) \cite{mooc,reich}. This increase in completion rate may be attributed in part to the selection of highly motivated students based on the admission criteria described above. 

Students completed open and closed-ended surveys after the course ended which we used in evaluating the course (n=151). We performed qualitative and quantitative analyses on these data. For the quantitative self-report data, students responded to 5-point Likert scale statements by choosing options from “strongly disagree” to “strongly agree”. In all cases, a p-value less than 0.05 was considered statistically significant. For the qualitative data, we read through the feedback and summarized common themes along with some quotes. 

\subsection{Proficiency in Programming Concepts}
\begin{figure}[htbp]
  \centering
  \includegraphics[width=0.5\linewidth]{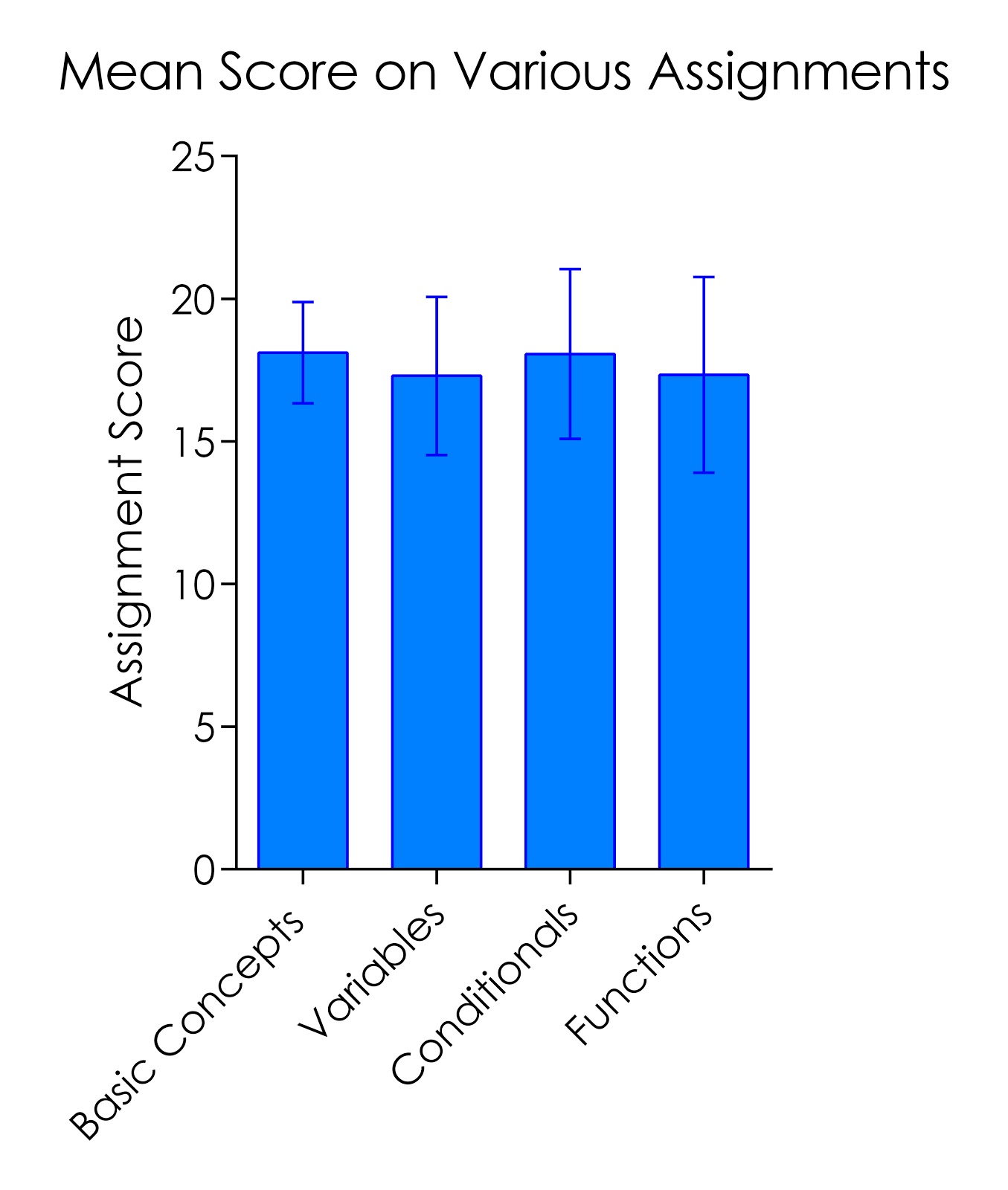}
  \caption{Average scores of students for each assignment. Bars represent mean±SD (n=151)}
  \Description{A bar chart}
\label{fig:avg_grades_fig}
\end{figure}

We evaluated the proficiency of students in the programming concepts that we taught them. We computed the average of the grades for each assignment (out of a maximum of 20 points per assignment) and show these in Figure \ref{fig:avg_grades_fig}. The result shows averages 17 and above, implying that students performed well on the assignments.

\begin{figure}[ht]
  \centering
  \includegraphics[width=0.8\linewidth]{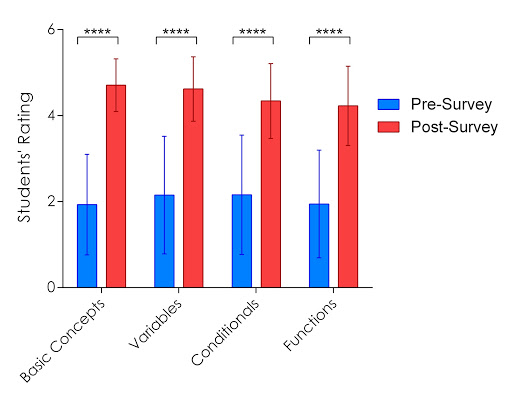}
  \caption{Comparison of students’ self-reported rating of understanding various programming concepts in pre and post-surveys. Bars represent Mean ±SD (n=151). The data was analyzed by a paired t-test. ****p<0.0001.}
  \Description{A bar chart}
\label{fig:understanding_programming_fig}
\end{figure}

We also evaluated students’ self-reported proficiency of the programming concepts covered in the course using a 5-point Likert scale (Table \ref{tab:concepts_survey}) in surveys before and after the course (Figure \ref{fig:understanding_programming_fig}). We performed paired t-tests to evaluate statistical significance. The results show that there was a statistically significant improvement in self-reported proficiency of the programming concepts which is consistent with the results of our past work \cite{boateng2018}. These results demonstrate that the SuaCode Africa course was able to improve students' proficiency in basic coding skills.

\begin{table}[ht]
\centering
\caption{Statements that students were asked to respond in order to assess their understanding of various programming concepts in pre and post surveys (1: Strongly Disagree to 5: Strongly Agree)}
\label{tab:concepts_survey}
\begin{tabular}{|l|l|}
\hline
\textbf{Concept} & \textbf{Statement} \\ \hline
Basic Concepts & I know basic concepts in Processing \\ \hline
Variables & I know how to create and use variables \\ \hline
Conditionals & I know how to create and use conditionals \\ \hline
Functions & I know how to create and use functions \\ \hline
\end{tabular}
\end{table}

\subsection{Understanding Lesson Notes and Assignment Difficulty} 
\begin{figure}[ht]
  \centering
  \includegraphics[width=0.8\linewidth]{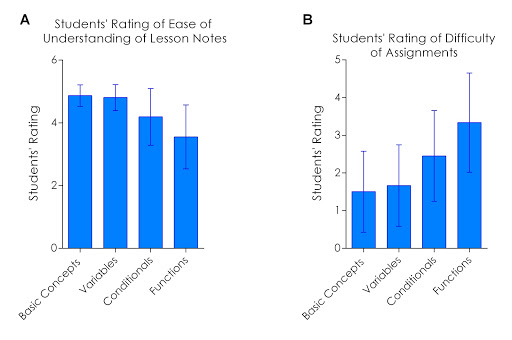}
  \caption{Students’ rating of the ease of understanding of lesson notes (A) and the difficulty of various assignments (B).}
  \Description{A bar chart}
\label{fig:lesson_assignment_difficulty}
\end{figure}

We also assessed how easy it was to understand our lesson notes. Using a 5-point Likert scale, students responded to the statement “the lesson was easy to understand” for each lesson from “strongly disagree” to “strongly agree”. Students also evaluated the difficulty of the assignments by responding to the statement “the assignment was difficult”. These results are summarised in Figure \ref{fig:lesson_assignment_difficulty}. Lesson notes were generally understandable for students with the lowest rating being 3.6±1.02 for the lesson on functions. These results points to a need to update the lesson note on functions to make it easier to understand. Overall, students mentioned in their free-response feedback on the course that the lesson notes were easy to understand. As expected, the rating of assignment difficulty increased as the course progressed through the lessons, with assignment on functions being the most challenging (3.4±1.32).

\subsection{Effects of Demographics on Performance}
\begin{figure}[ht]
  \centering
  \includegraphics[width=0.8\linewidth]{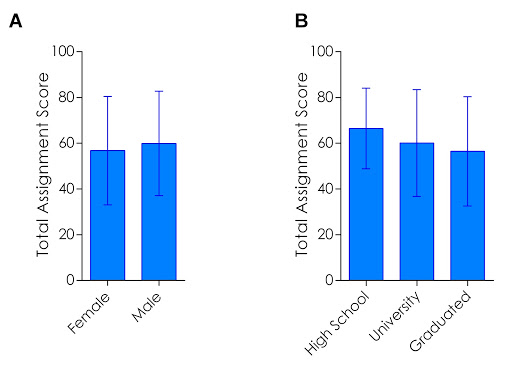}
  \caption{Total Assignment score of students grouped by gender(A) and educational level (B). Data represent the mean±SD}
  \Description{A bar chart}
\label{fig:grades_by_education_and_gender}
\end{figure}

Since we had a large and diverse cohort, we decided to evaluate whether there was a difference in course performance for different demographic groups of students. First, we assessed performance by gender. The completion rates for males and females are shown in the contingency Table \ref{tab:completion_gender}. Analysis of this data by Chi-square test revealed a non-significant association between gender and completion of the course (p=0.382). This corroborates previous studies that indicate that across domains, female students tend to complete online courses at the same rate as male students \cite{breslow2013, cisel2014}. Furthermore, the total assignment scores of females (56.7±23.7) and males (59.9±22.8) compared by an unpaired t-test revealed non-significant differences (p=0.414) (Figure \ref{fig:grades_by_education_and_gender}). According to the EQUALS report, Africa has the lowest percentage of women employees in the IT field, only 18\%, and this is due to several factors including inadequate training opportunities and access to digital infrastructure \cite{sey2019}. Thus it is interesting to note the females performed comparably to males in our course. Considering the potential of scaling this program, it offers a viable solution to bridging the gender divide in coding education.

\begin{table}[ht]
\centering
\caption{Completion rates analyzed by gender}
\label{tab:completion_gender}
\begin{tabular}{|p{0.12\linewidth}|p{0.2\linewidth}|p{0.3\linewidth}|p{0.17\linewidth}|}
\hline
\textbf{Gender} & \textbf{Completed} & \textbf{Did Not Complete} & \textbf{Total(\%)} \\ \hline
Female & 31 (14.8\%) & 9 (4.3\%) & 40 (19.1\%) \\ \hline
Male & 120 (57.1\%) & 50 (23.8\%) & 170 (80.9\%) \\ \hline
Total & 151 (71.9\%) & 59 (28.1\%) & 210 (100\%) \\ \hline
\end{tabular}
\end{table}

We also analyzed completion rates across educational levels. The data are shown in the contingency Table \ref{tab:completion_education}. Four students (3 primary school students, 1 student taking a gap year after high school) were excluded from this analysis because the numbers were too few to create new categories. Interestingly, analysis by a Chi-square test revealed a non-significant association between the current educational level and completion of the course (p=0.149). Furthermore, the total assignment scores of high school students, university students, and university graduates were compared by one-way analysis of variance (ANOVA) test (Figure \ref{fig:grades_by_education_and_gender}). There was no significant difference detected between the groups  (\(F_2, _{203}\)=1.886; p=0.1543). These results indicate that performance and completion is comparable for all educational level of students. Hence, the course could be suitable for the introduction of coding to students across educational levels.

\begin{table}[ht]
\centering
\caption{Completion rates analyzed by current education level}
\label{tab:completion_education}
\begin{tabular}{|p{0.268\linewidth}|p{0.175\linewidth}|p{0.299\linewidth}|p{0.163\linewidth}|}
\hline
\textbf{Education Level} & \textbf{Completed} & \textbf{Did Not Complete} & \textbf{Total} \\ \hline
High School & 20 (9.7\%) & 3 (1.5\%) & 23 (11.2\%) \\ \hline
University & 63 (30.6\%) & 23 (11.2\%) & 86 (41.7\%) \\ \hline
Graduated & 65 (31.6\%) & 32 (15.5\%) & 97 (47.1\%) \\ \hline
Total & 148 (71.8\%) & 58 (28.2\%) & 206 (100\%) \\ \hline
\end{tabular}
\end{table}

\subsection{Smartphone Coding Experience}
\begin{figure}[ht]
  \centering
  \includegraphics[width=0.8\linewidth]{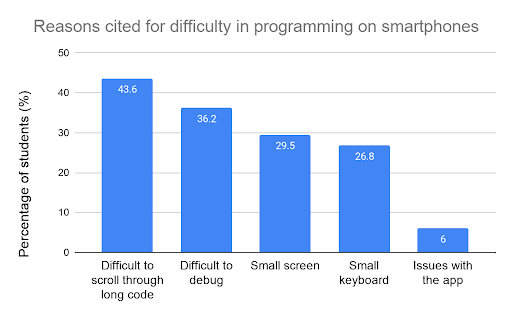}
  \caption{Top five reasons cited for the difficulty in programming on smartphones with multiple selections allowed}
  \Description{A bar chart}
\label{fig:smartphone_coding_challenge}
\end{figure}

We also assessed the smartphone-coding experience of students. When asked to rate their agreement to the statement “I enjoyed programming using my smartphone” on a scale ranging from 1 (strongly disagree) to 5 (strongly agree), the mean response was 4.5±0.78. To the statement, “I would like to code more on my smartphone”, the mean was 4.2±1.15. These results are favorable and support the belief that smartphones are a good medium to introduce coding skills in Africa. When asked what they thought about coding on a smartphone, the feedback was mostly positive and the experience was often described as “fun, easy and accessible”. We hypothesize that the nature of our exercises and assignments may have contributed to the enjoying experience of coding on a smartphone since we used a game-based approach which has been shown to be an interesting, engaging and motivating way of introducing novices to coding \cite{leutenegger2007games,Barnes:2008,Bayliss:2006} and more so for mobile games \cite{Kurkovsky:2009}. Many students did not own a laptop and appreciated the convenience of working on the device they were most used to. However, some participants complained about the complications arising from the relatively small size of the screen. The main challenges to programming on smartphones cited by students are shown in Figure \ref{fig:smartphone_coding_challenge} (students could select multiple options). These findings are in agreement with our past works \cite{boateng2018, boateng2019}. However, it is interesting to note that these challenges did not impede them obtaining proficiency in coding as shown in a previous section.

\subsection{Peer and Facilitator Assistance}
The teaching pedagogy of SuaCode Africa followed a student-centered approach. We evaluated if students received adequate assistance from their peers and facilitators. In their feedback, students mentioned they appreciated the help they got. For example, one student said, “The facilitators and even colleagues were always available to help” in response to the question "What suggestions do you have for improving the help provided?". Students responded to a 5-point Likert scale on facilitators’ helpfulness and timeliness. Overall, the average rating of 4.67±0.66 and 4.25±0.88 for helpfulness and timeliness respectively shows that students were satisfied with the assistance they received from peers and facilitators in the course forum. In addition to using the course forum, students also created groups on WhatsApp (which is popular across Africa) to answer each other’s questions which shows that there was an interest to help each other. Using a cohort-based approach may have contributed to this sense of community within SuaCode Africa since students were all working with the same timeline and cohort-based learning fulfills students’ need for affiliation, support, interaction, and teamwork \cite{maher2005}.

\subsection{Mentoring Session}
After the course ended, we organized a 1-hour online video mentoring session in which the top students (those who had perfect scores on all assignments) were invited to have a conversation with a Software Engineer at Google. During the session, the engineer shared his life story on getting into programming and answered various questions from students on how to advance and build their coding skills. Students mentioned that they enjoyed the session and appreciated the responses to their questions.

\subsection{Impact on Students}
The students that completed SuaCode Africa were sent a survey a year later that asked them to share a short paragraph on what they had been up to since the program with reference to building upon the skills from SuaCode Africa. These surveys were sent via the emails they used during the program with 21 students responding. Students reported that they had been involved in active software development and that SuaCode  Africa was instrumental in developing the foundation that led to their advancement. Fifteen of them had learned more than one programming language and two had written programming tutorials. One of them had built a payment processing system for a local restaurant and three of them had led in-person training workshops in Augmented Reality, Fullstack development, and microcontroller programming. Some of them reported reviewing the lessons, improving the software they wrote during the SuaCode program, and adding new features. Some languages and tools learned since the SuaCode program included Python, HTML, CSS, C++, C\#, JavaScript, Java, PHP, SoldWorks, Processing, Internet of Things, and Flutter. Lastly, one student wrote an article about his experience in the SuaCode program and published the lessons and the Pong game he built in a Medium article \cite{medium_article1}. Overall, SuaCode Africa laid the foundation for these students to build further coding skills.

\section{Challenges and Future Work}
We had a challenge with accommodating all 709 students in Google Classroom since the free version only accommodates a maximum of 250 people. Hence, we created 3 different classrooms to accommodate all the students. This situation resulted in posting lesson materials and assignments three times and also answering the same questions asked in the three classrooms. In the future, we plan to use a platform that can accommodate a larger number like Piazza.

We had Francophone students in the course who struggled to understand the course materials which were in English. They also sometimes posted questions in French which we did not plan for. Given that French is a major official language in countries in Africa, in the future, we will provide the course materials in French and also support and answer questions asked in French so that language is not a barrier to learning to code among African students.

Some students also repeatedly asked questions that had answers in the lesson documents and under previous posts in the forum. A number of limitations of the Google Classroom platform contributed to this problem. Firstly, the mobile interface of the Classroom's forum displayed the whole contents of questions and so students had to scroll a lot to find previous questions. Additionally, there was no option to search for or filter previous questions. Hence, students tended to post questions without finding out if it had already been asked. This situation might have contributed to students creating and using WhatsApp groups and also increased the question-answering burden on facilitators. To address these issues, we plan to use a platform that facilitates easy searching and filtering of past questions like Piazza. Additionally, we plan to develop an AI-based question answering system such as by Goel \cite{goel2020} that will automatically answer the questions of students and reduce the burden on facilitators.

\section{Conclusion}
In this work, we ran SuaCode Africa which received applications from 709 Africans in 37 countries. Out of the 210 students admitted to the course, 72\% completed. Additionally, students’ assignment submissions and self-reports showed proficiency in the programming concepts we taught, with comparable performance between males and females, and across educational levels. Also, students mentioned that the lesson notes were easy to understand, and they enjoyed the experience of writing code on their smartphones. Furthermore, the assistance provided by peers and facilitators on the course forum was adequate for students. A year after completion, some students were still engaged in coding by developing various applications, writing online tutorials, reviewing and improving their apps, and pursuing further learning opportunities. In conclusion, SuaCode Africa was effective at introducing students across Africa to coding using smartphones and inspiring an interest in further digital skills acquisition. We are excited about bringing coding skills to the fingertips of millions across Africa through SuaCode.

\begin{acks}
We are grateful to the Processing Foundation for supporting this work via the Processing Foundation Fellowship program. Also, we thank the SuaCode team that assisted in running the program. 
\end{acks}

\balance

\bibliographystyle{ACM-Reference-Format}
\bibliography{ref}


\begin{thebibliography}{31}


\ifx \showCODEN    \undefined \def \showCODEN     #1{\unskip}     \fi
\ifx \showDOI      \undefined \def \showDOI       #1{#1}\fi
\ifx \showISBNx    \undefined \def \showISBNx     #1{\unskip}     \fi
\ifx \showISBNxiii \undefined \def \showISBNxiii  #1{\unskip}     \fi
\ifx \showISSN     \undefined \def \showISSN      #1{\unskip}     \fi
\ifx \showLCCN     \undefined \def \showLCCN      #1{\unskip}     \fi
\ifx \shownote     \undefined \def \shownote      #1{#1}          \fi
\ifx \showarticletitle \undefined \def \showarticletitle #1{#1}   \fi
\ifx \showURL      \undefined \def \showURL       {\relax}        \fi
\providecommand\bibfield[2]{#2}
\providecommand\bibinfo[2]{#2}
\providecommand\natexlab[1]{#1}
\providecommand\showeprint[2][]{arXiv:#2}

\bibitem[\protect\citeauthoryear{??}{acw}{[n.d.]}]%
        {acw}
 \bibinfo{year}{[n.d.]}\natexlab{}.
\newblock \bibinfo{title}{Africa Code Week}.
\newblock
\newblock
\urldef\tempurl%
\url{https://africacodeweek.org/}
\showURL{%
Retrieved July, 2020 from \tempurl}


\bibitem[\protect\citeauthoryear{??}{Ade}{2007}]%
        {Aderinoye_Ojokheta_Olojede_2007}
 \bibinfo{year}{2007}\natexlab{}.
\newblock \showarticletitle{Integrating Mobile Learning into Nomadic Education
  Programme in Nigeria: Issues and perspectives}.
\newblock   \bibinfo{volume}{8} (\bibinfo{date}{Jun.} \bibinfo{year}{2007}).
\newblock
\urldef\tempurl%
\url{https://doi.org/10.19173/irrodl.v8i2.347}
\showDOI{\tempurl}


\bibitem[\protect\citeauthoryear{??}{sua}{2018}]%
        {suacode-course}
 \bibinfo{year}{2018}\natexlab{}.
\newblock \bibinfo{title}{Suacode - Smartphone-Based Coding Course}.
\newblock
\newblock
\urldef\tempurl%
\url{https://github.com/Suacode-app/Suacode}
\showURL{%
Retrieved July, 2020 from \tempurl}


\bibitem[\protect\citeauthoryear{??}{aut}{2019}]%
        {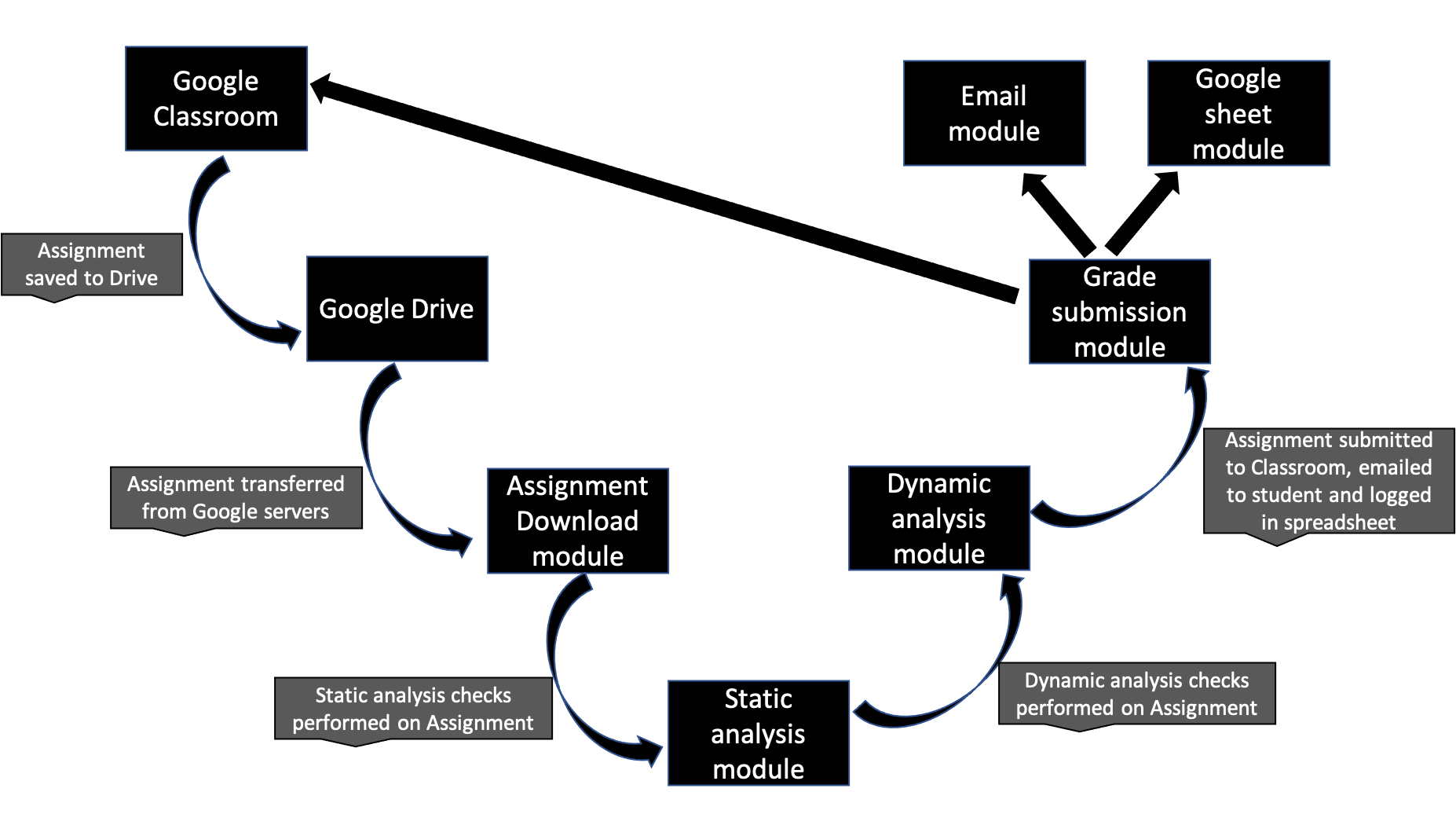}
 \bibinfo{year}{2019}\natexlab{}.
\newblock \bibinfo{title}{AutoGrad: Automatic grading system for SuaCode}.
\newblock
\newblock
\urldef\tempurl%
\url{https://github.com/PrinceSAnnor/autograd}
\showURL{%
Retrieved July, 2020 from \tempurl}


\bibitem[\protect\citeauthoryear{Africa}{Africa}{2019}]%
        {QuartzAfrica:2019}
\bibfield{author}{\bibinfo{person}{Quartz Africa}.}
  \bibinfo{year}{2019}\natexlab{}.
\newblock \bibinfo{booktitle}{\emph{The cost of internet access is dropping
  globally but not fast enough in Africa}}.
\newblock
\urldef\tempurl%
\url{https://qz.com/africa/1577429/how-much-is-1gb-of-mobile-data-in-africa/}
\showURL{%
Retrieved June, 2019 from \tempurl}


\bibitem[\protect\citeauthoryear{APDE}{APDE}{[n.d.]}]%
        {apde}
APDE \bibinfo{year}{[n.d.]}\natexlab{}.
\newblock \bibinfo{booktitle}{\emph{APDE - Android Processing IDE}}.
\newblock
\urldef\tempurl%
\url{https://play.google.com/store/apps/details?id=com.calsignlabs.apde}
\showURL{%
Retrieved Jan, 2019 from \tempurl}


\bibitem[\protect\citeauthoryear{Barnes, Powell, Chaffin, and Lipford}{Barnes
  et~al\mbox{.}}{2008}]%
        {Barnes:2008}
\bibfield{author}{\bibinfo{person}{Tiffany Barnes}, \bibinfo{person}{Eve
  Powell}, \bibinfo{person}{Amanda Chaffin}, {and} \bibinfo{person}{Heather
  Lipford}.} \bibinfo{year}{2008}\natexlab{}.
\newblock \showarticletitle{Game2Learn: Improving the Motivation of CS1
  Students}. In \bibinfo{booktitle}{\emph{Proceedings of the 3rd International
  Conference on Game Development in Computer Science Education}} (Miami,
  Florida) \emph{(\bibinfo{series}{GDCSE '08})}. \bibinfo{publisher}{ACM},
  \bibinfo{address}{New York, NY, USA}, \bibinfo{pages}{1--5}.
\newblock
\showISBNx{978-1-60558-057-9}
\urldef\tempurl%
\url{https://doi.org/10.1145/1463673.1463674}
\showDOI{\tempurl}


\bibitem[\protect\citeauthoryear{Bayliss and Strout}{Bayliss and
  Strout}{2006}]%
        {Bayliss:2006}
\bibfield{author}{\bibinfo{person}{Jessica~D. Bayliss} {and}
  \bibinfo{person}{Sean Strout}.} \bibinfo{year}{2006}\natexlab{}.
\newblock \showarticletitle{Games As a "Flavor" of CS1}. In
  \bibinfo{booktitle}{\emph{Proceedings of the 37th SIGCSE Technical Symposium
  on Computer Science Education}} (Houston, Texas, USA)
  \emph{(\bibinfo{series}{SIGCSE '06})}. \bibinfo{publisher}{ACM},
  \bibinfo{address}{New York, NY, USA}, \bibinfo{pages}{500--504}.
\newblock
\showISBNx{1-59593-259-3}
\urldef\tempurl%
\url{https://doi.org/10.1145/1121341.1121498}
\showDOI{\tempurl}


\bibitem[\protect\citeauthoryear{Boateng and Kumbol}{Boateng and
  Kumbol}{2018}]%
        {boateng2018}
\bibfield{author}{\bibinfo{person}{George Boateng} {and}
  \bibinfo{person}{Victor Kumbol}.} \bibinfo{year}{2018}\natexlab{}.
\newblock \showarticletitle{Project iSWEST: Promoting a culture of innovation
  in Africa through STEM}. In \bibinfo{booktitle}{\emph{2018 IEEE Integrated
  STEM Education Conference (ISEC)}}. \bibinfo{pages}{104--111}.
\newblock
\urldef\tempurl%
\url{https://doi.org/10.1109/ISECon.2018.8340459}
\showDOI{\tempurl}


\bibitem[\protect\citeauthoryear{Boateng, Kumbol, and Annor}{Boateng
  et~al\mbox{.}}{2019}]%
        {boateng2019}
\bibfield{author}{\bibinfo{person}{George Boateng}, \bibinfo{person}{Victor
  Wumbor-Apin Kumbol}, {and} \bibinfo{person}{Prince~Steven Annor}.}
  \bibinfo{year}{2019}\natexlab{}.
\newblock \showarticletitle{Keep Calm and Code on Your Phone: A Pilot of
  SuaCode, an Online Smartphone-Based Coding Course}. In
  \bibinfo{booktitle}{\emph{Proceedings of the 8th Computer Science Education
  Research Conference}}. \bibinfo{pages}{9--14}.
\newblock


\bibitem[\protect\citeauthoryear{Breslow, Pritchard, DeBoer, Stump, Ho, and
  Seaton}{Breslow et~al\mbox{.}}{2013}]%
        {breslow2013}
\bibfield{author}{\bibinfo{person}{Lori Breslow}, \bibinfo{person}{David~E
  Pritchard}, \bibinfo{person}{Jennifer DeBoer}, \bibinfo{person}{Glenda~S
  Stump}, \bibinfo{person}{Andrew~D Ho}, {and} \bibinfo{person}{Daniel~T
  Seaton}.} \bibinfo{year}{2013}\natexlab{}.
\newblock \showarticletitle{Studying learning in the worldwide classroom
  research into edX's first MOOC.}
\newblock \bibinfo{journal}{\emph{Research \& Practice in Assessment}}
  \bibinfo{volume}{8} (\bibinfo{year}{2013}), \bibinfo{pages}{13--25}.
\newblock


\bibitem[\protect\citeauthoryear{Cisel}{Cisel}{2014}]%
        {cisel2014}
\bibfield{author}{\bibinfo{person}{Matthieu Cisel}.}
  \bibinfo{year}{2014}\natexlab{}.
\newblock \showarticletitle{Analyzing completion rates in the First French
  xMOOC}.
\newblock \bibinfo{journal}{\emph{Proceedings of the European MOOC stakeholder
  summit}}  \bibinfo{volume}{26} (\bibinfo{year}{2014}), \bibinfo{pages}{51}.
\newblock


\bibitem[\protect\citeauthoryear{Elkhateeb, Shehab, and El-Bakry}{Elkhateeb
  et~al\mbox{.}}{2019}]%
        {elkhateeb2019mobile}
\bibfield{author}{\bibinfo{person}{Menna Elkhateeb}, \bibinfo{person}{Abdulaziz
  Shehab}, {and} \bibinfo{person}{Hazem El-Bakry}.}
  \bibinfo{year}{2019}\natexlab{}.
\newblock \showarticletitle{Mobile learning system for egyptian higher
  education using agile-based approach}.
\newblock \bibinfo{journal}{\emph{Education Research International}}
  \bibinfo{volume}{2019} (\bibinfo{year}{2019}).
\newblock


\bibitem[\protect\citeauthoryear{Goel}{Goel}{2020}]%
        {goel2020}
\bibfield{author}{\bibinfo{person}{Ashok Goel}.}
  \bibinfo{year}{2020}\natexlab{}.
\newblock \showarticletitle{AI-Powered Learning: Making Education Accessible,
  Affordable, and Achievable}.
\newblock \bibinfo{journal}{\emph{arXiv preprint arXiv:2006.01908}}
  (\bibinfo{year}{2020}).
\newblock


\bibitem[\protect\citeauthoryear{Ibeabuchi}{Ibeabuchi}{2019}]%
        {medium_article1}
\bibfield{author}{\bibinfo{person}{Onyedikachi~E. Ibeabuchi}.}
  \bibinfo{year}{2019}\natexlab{}.
\newblock \bibinfo{title}{Create a Pong Game Using an Android Device}.
\newblock
\newblock
\urldef\tempurl%
\url{https://medium.com/@eonyedikachi/create-pong-game-using-an-android-phone-b782579ee950}
\showURL{%
Retrieved July, 2020 from \tempurl}


\bibitem[\protect\citeauthoryear{Jacob and Issac}{Jacob and Issac}{2007}]%
        {Jacob2007MobileLI}
\bibfield{author}{\bibinfo{person}{S.~M. Jacob} {and} \bibinfo{person}{B.
  Issac}.} \bibinfo{year}{2007}\natexlab{}.
\newblock \showarticletitle{Mobile learning in transforming higher education}.
\newblock


\bibitem[\protect\citeauthoryear{Kafyulilo}{Kafyulilo}{2014}]%
        {Ayoub}
\bibfield{author}{\bibinfo{person}{Ayoub Kafyulilo}.}
  \bibinfo{year}{2014}\natexlab{}.
\newblock \showarticletitle{Access, Use and Perceptions of Teachers and
  Students towards Mobile Phones as a Tool for Teaching and Learning in
  Tanzania}.
\newblock \bibinfo{journal}{\emph{Education and Information Technologies}}
  \bibinfo{volume}{19}, \bibinfo{number}{1} (\bibinfo{date}{March}
  \bibinfo{year}{2014}), \bibinfo{pages}{115–127}.
\newblock
\showISSN{1360-2357}
\urldef\tempurl%
\url{https://doi.org/10.1007/s10639-012-9207-y}
\showDOI{\tempurl}


\bibitem[\protect\citeauthoryear{Kizilcec and Halawa}{Kizilcec and
  Halawa}{2015}]%
        {mooc}
\bibfield{author}{\bibinfo{person}{Ren\'{e}~F. Kizilcec} {and}
  \bibinfo{person}{Sherif Halawa}.} \bibinfo{year}{2015}\natexlab{}.
\newblock \showarticletitle{Attrition and Achievement Gaps in Online Learning}.
  In \bibinfo{booktitle}{\emph{Proceedings of the Second (2015) ACM Conference
  on Learning @ Scale}} (Vancouver, BC, Canada) \emph{(\bibinfo{series}{L@S
  '15})}. \bibinfo{publisher}{Association for Computing Machinery},
  \bibinfo{address}{New York, NY, USA}, \bibinfo{pages}{57–66}.
\newblock
\showISBNx{9781450334112}
\urldef\tempurl%
\url{https://doi.org/10.1145/2724660.2724680}
\showDOI{\tempurl}


\bibitem[\protect\citeauthoryear{Kukulska-Hulme, Sharples, Milrad,
  Arnedillo-S{\'a}nchez, and Vavoula}{Kukulska-Hulme et~al\mbox{.}}{2009}]%
        {Kukulska}
\bibfield{author}{\bibinfo{person}{Agnes Kukulska-Hulme}, \bibinfo{person}{Mike
  Sharples}, \bibinfo{person}{Marcelo Milrad}, \bibinfo{person}{Inmaculada
  Arnedillo-S{\'a}nchez}, {and} \bibinfo{person}{Giasemi Vavoula}.}
  \bibinfo{year}{2009}\natexlab{}.
\newblock \showarticletitle{Innovation in Mobile Learning: A European
  Perspective}.
\newblock \bibinfo{journal}{\emph{International Journal of Mobile and Blended
  Learning}} \bibinfo{volume}{1}, \bibinfo{number}{1} (\bibinfo{date}{January}
  \bibinfo{year}{2009}), \bibinfo{pages}{13--35}.
\newblock
\urldef\tempurl%
\url{http://oro.open.ac.uk/12711/}
\showURL{%
\tempurl}


\bibitem[\protect\citeauthoryear{Kurkovsky}{Kurkovsky}{2009}]%
        {Kurkovsky:2009}
\bibfield{author}{\bibinfo{person}{Stan Kurkovsky}.}
  \bibinfo{year}{2009}\natexlab{}.
\newblock \showarticletitle{Engaging Students Through Mobile Game Development}.
  In \bibinfo{booktitle}{\emph{Proceedings of the 40th ACM Technical Symposium
  on Computer Science Education}} (Chattanooga, TN, USA)
  \emph{(\bibinfo{series}{SIGCSE '09})}. \bibinfo{publisher}{ACM},
  \bibinfo{address}{New York, NY, USA}, \bibinfo{pages}{44--48}.
\newblock
\showISBNx{978-1-60558-183-5}
\urldef\tempurl%
\url{https://doi.org/10.1145/1508865.1508881}
\showDOI{\tempurl}


\bibitem[\protect\citeauthoryear{Leutenegger and Edgington}{Leutenegger and
  Edgington}{2007}]%
        {leutenegger2007games}
\bibfield{author}{\bibinfo{person}{Scott Leutenegger} {and}
  \bibinfo{person}{Jeffrey Edgington}.} \bibinfo{year}{2007}\natexlab{}.
\newblock \showarticletitle{A games first approach to teaching introductory
  programming}. In \bibinfo{booktitle}{\emph{ACM SiGCSE Bulletin}},
  Vol.~\bibinfo{volume}{39}. ACM, \bibinfo{pages}{115--118}.
\newblock


\bibitem[\protect\citeauthoryear{Maher}{Maher}{2005}]%
        {maher2005}
\bibfield{author}{\bibinfo{person}{Michelle~A Maher}.}
  \bibinfo{year}{2005}\natexlab{}.
\newblock \showarticletitle{The evolving meaning and influence of cohort
  membership}.
\newblock \bibinfo{journal}{\emph{Innovative Higher Education}}
  \bibinfo{volume}{30}, \bibinfo{number}{3} (\bibinfo{year}{2005}),
  \bibinfo{pages}{195--211}.
\newblock


\bibitem[\protect\citeauthoryear{Maleko, Hamilton, and D'Souza}{Maleko
  et~al\mbox{.}}{2012}]%
        {maleko}
\bibfield{author}{\bibinfo{person}{Mercy Maleko}, \bibinfo{person}{Margaret
  Hamilton}, {and} \bibinfo{person}{Daryl D'Souza}.}
  \bibinfo{year}{2012}\natexlab{}.
\newblock \showarticletitle{Novices' Perceptions and Experiences of a Mobile
  Social Learning Environment for Learning of Programming}. In
  \bibinfo{booktitle}{\emph{Proceedings of the 17th ACM Annual Conference on
  Innovation and Technology in Computer Science Education}} (Haifa, Israel)
  \emph{(\bibinfo{series}{ITiCSE '12})}. \bibinfo{publisher}{Association for
  Computing Machinery}, \bibinfo{address}{New York, NY, USA},
  \bibinfo{pages}{285–290}.
\newblock
\showISBNx{9781450312462}
\urldef\tempurl%
\url{https://doi.org/10.1145/2325296.2325364}
\showDOI{\tempurl}


\bibitem[\protect\citeauthoryear{Matinde}{Matinde}{2016}]%
        {matinde2017}
\bibfield{author}{\bibinfo{person}{Vincent Matinde}.}
  \bibinfo{year}{2016}\natexlab{}.
\newblock \bibinfo{booktitle}{\emph{Africa 2017: Smartphone penetration, Open
  Data and less online freedom,}}.
\newblock
\urldef\tempurl%
\url{https://www.idgconnect.com/idgconnect/opinion/1022805/africa-2017-smartphone-penetration-online-freedom}
\showURL{%
Retrieved Jan, 2019 from \tempurl}


\bibitem[\protect\citeauthoryear{Mbogo, Blake, and Suleman}{Mbogo
  et~al\mbox{.}}{2014}]%
        {mbogo2014supporting}
\bibfield{author}{\bibinfo{person}{Chao Mbogo}, \bibinfo{person}{Edwin Blake},
  {and} \bibinfo{person}{Hussein Suleman}.} \bibinfo{year}{2014}\natexlab{}.
\newblock \showarticletitle{Supporting the Construction of Programs on a Mobile
  Device: A Scaffolding Framework}.
\newblock  (\bibinfo{year}{2014}).
\newblock


\bibitem[\protect\citeauthoryear{Mbogo, Blake, and Suleman}{Mbogo
  et~al\mbox{.}}{2016}]%
        {mbogo2016evaluating}
\bibfield{author}{\bibinfo{person}{Chao Mbogo}, \bibinfo{person}{Edwin Blake},
  {and} \bibinfo{person}{Hussein Suleman}.} \bibinfo{year}{2016}\natexlab{}.
\newblock \showarticletitle{Evaluating the effect of using scaffolding
  techniques to support Java programming on a mobile phone}.
\newblock \bibinfo{journal}{\emph{IADIS International Journal on WWW/Internet}}
  \bibinfo{volume}{14}, \bibinfo{number}{1} (\bibinfo{year}{2016}).
\newblock


\bibitem[\protect\citeauthoryear{Processing}{Processing}{[n.d.]}]%
        {processing}
Processing \bibinfo{year}{[n.d.]}\natexlab{}.
\newblock \bibinfo{booktitle}{\emph{Processing Foundation}}.
\newblock
\urldef\tempurl%
\url{https://processing.org/}
\showURL{%
Retrieved Jan, 2019 from \tempurl}


\bibitem[\protect\citeauthoryear{Reich}{Reich}{[n.d.]}]%
        {reich}
\bibfield{author}{\bibinfo{person}{Justin Reich}.}
  \bibinfo{year}{[n.d.]}\natexlab{}.
\newblock \bibinfo{title}{MOOC Completion and Retention in the Context of
  Student Intent}.
\newblock
\newblock
\urldef\tempurl%
\url{https://er.educause.edu/articles/2014/12/mooc-completion-and-retention-in-the-context-of-student-intent}
\showURL{%
\tempurl}


\bibitem[\protect\citeauthoryear{SAP}{SAP}{2016}]%
        {sap2016}
\bibfield{author}{\bibinfo{person}{SAP}.} \bibinfo{year}{2016}\natexlab{}.
\newblock \bibinfo{title}{Africa Code Week - Bridging the Digital Skills Gap in
  Africa}.
\newblock
\newblock
\urldef\tempurl%
\url{https://africacodeweek.org/fr/blog/https-www.linkedin.com-pulse-africa-code-week-bridging-digital-skills-gap}
\showURL{%
Retrieved July, 2020 from \tempurl}


\bibitem[\protect\citeauthoryear{Sey and Hafkin}{Sey and Hafkin}{2019}]%
        {sey2019}
\bibfield{author}{\bibinfo{person}{Araba Sey} {and} \bibinfo{person}{Nancy
  Hafkin}.} \bibinfo{year}{2019}\natexlab{}.
\newblock \showarticletitle{Taking stock: Data and evidence on gender equality
  in digital access, skills and leadership}.
\newblock \bibinfo{journal}{\emph{United Nations University, Tokyo}}
  (\bibinfo{year}{2019}).
\newblock


\bibitem[\protect\citeauthoryear{Shonola, Joy, Oyelere, and Suhonen}{Shonola
  et~al\mbox{.}}{2016}]%
        {Shonola2016TheIO}
\bibfield{author}{\bibinfo{person}{S.~A. Shonola}, \bibinfo{person}{M. Joy},
  \bibinfo{person}{S. Oyelere}, {and} \bibinfo{person}{Jarkko Suhonen}.}
  \bibinfo{year}{2016}\natexlab{}.
\newblock \showarticletitle{The Impact of Mobile Devices for Learning in Higher
  Education Institutions: Nigerian Universities Case Study}.
\newblock \bibinfo{journal}{\emph{International Journal of Modern Education and
  Computer Science}}  \bibinfo{volume}{8} (\bibinfo{year}{2016}),
  \bibinfo{pages}{43--50}.
\newblock


\end{thebibliography}

\end{document}